\documentclass[
superscriptaddress,
 amsmath,amssymb,
 aps,
reprint
]{revtex4-2}

\usepackage{color}
\usepackage{graphicx}
\usepackage{dcolumn}
\usepackage{bm}
\usepackage{epstopdf}
\usepackage{gensymb}

\usepackage{float}

\begin{document}

\title{Control of ternary alloy composition during remote epitaxy on graphene}

\author{Zachary LaDuca}
\affiliation{Materials Science and Engineering, University of Wisconsin-Madison, Madison, WI 53706}

\author{Katherine Su}
\affiliation{Materials Science and Engineering, University of Wisconsin-Madison, Madison, WI 53706}

\author{Sebastian Manzo}
\affiliation{Materials Science and Engineering, University of Wisconsin-Madison, Madison, WI 53706}

\author{Michael S. Arnold}
\affiliation{Materials Science and Engineering, University of Wisconsin-Madison, Madison, WI 53706}

\author{Jason K. Kawasaki}
\affiliation{Materials Science and Engineering, University of Wisconsin-Madison, Madison, WI 53706}
\email{jkawasaki@wisc.edu}

\date{\today}
\begin{abstract}

Understanding the sticking coefficient $\sigma$, i.e., the probability of an adatom sticking to a surface, is essential for controlling the stoichiometry during epitaxial film growth. However, $\sigma$ on monolayer graphene-covered surfaces and its impact on remote epitaxy are not understood. Here, using molecular-beam epitaxial (MBE) growth of the magnetic shape memory alloy Ni$_2$MnGa, we show that the sticking coefficients for metals on graphene-covered MgO (001) are less than one and are temperature and element dependent, as revealed by ion backscattering spectrometry (IBS) and energy dispersive x-ray spectroscopy (EDS). This lies in stark contrast with most transition metals sticking on semiconductor and oxide substrates, for which $\sigma$ is near unity at typical growth temperatures ($T<800\degree$C). By initiating growth below $400 \degree$ C, where the sticking coefficients are closer to unity and wetting on the graphene surface is improved, we demonstrate epitaxy of Ni$_2$MnGa films with controlled stoichiometry that can be exfoliated to produce freestanding membranes. Straining these membranes tunes the magnetic coercive field. Our results provide a route to synthesize membranes with complex stoichiometries whose properties can be manipulated via strain.

\end{abstract}

\maketitle

Remote \cite{kim2017remote, yoon2022freestanding} and van der Waals \cite{koma1992van, ren2021van, du2022controlling, kim2014principle} epitaxy on monolayer graphene-covered substrates are promising strategies for synthesizing single crystalline films that are mechanically decoupled from the substrate. In remote epitaxy, films are thought to grow on graphene-covered substrates with epitaxial registry to the substrate, due to the ``remote'' lattice potential of the substrate potential that permeates through graphene \cite{kim2017remote,kong2018polarity}. Applications include lattice mismatched epitaxy with reduced dislocation densities \cite{bae2020graphene, liu2022atomic}, etch-free exfoliation of membranes for flexible electronics and re-use of substrates \cite{kim2017remote}, and discovery of new properties induced by extreme strain and strain gradients in membranes \cite{du2021epitaxy, du2023strain, kum2020heterogeneous}. 

A fundamental challenge, however, is controlling the film stoichiometry during growth on graphene. Due to the weak van der Waals interactions, the sticking coefficients $\sigma$ for metals on multilayer graphite are typically $\sigma < 0.1$ at room temperature as measured by desorption spectroscopy \cite{arthur1973adsorption}, x-ray photoemission spectroscopy (XPS) \cite{lopez2005ag, howells2002effects}, and scanning tunneling microscopy (STM) \cite{lopez2005ag, howells2002effects}. This lies in stark contrast with the typical $\sigma \sim 1$ for metals on semiconductor, oxide, and metal surfaces \cite{appy2014transition}, which enables a simple one-to-one correspondence between film stoichiometry and incident flux ratios. Although $\sigma$ on monolayer graphene-covered surfaces is anticipated to be closer to unity \cite{appy2014transition}, due to the ``remote'' substrate interactions that permeate through graphene \cite{kim2017remote, rafiee2012wetting,kong2018polarity}, sticking on graphene is less understood and unlikely to be exactly 1. Moreover, the ``remote'' argument suggests that $\sigma$ on graphene-covered substrates should depend on the identity of the substrate. The impact of element-dependent, non unity sticking coefficients during growth on monolayer graphene is generally overlooked, in part, because remote epitaxy has focused on compound semiconductors like GaAs for which the stoichiometry is self-limited by growth within an adsorption-controlled window \cite{arthur1968interaction, cho1975molecular}. But for more complex materials like ternary transition metal oxides or intermetallic Heusler compounds adsorption-controlled growth windows are only accessible in select cases \cite{jalan2009molecular, yoon2022freestanding, ihlefeld2009adsorption, theis1998adsorption, shourov2020semi}. Controlling the stoichiometry of these materials during remote epitaxy or van der Waals epitaxy on graphene \cite{kum2020heterogeneous,yoon2022freestanding,ma2021remote,jia2021van,dai2022highly,du2021epitaxy,du2022controlling} in the ultrathin limit will require understanding the sticking coefficients on graphene.

Here, using MBE growth of the magnetic shape memory alloy Ni$_2$MnGa, we show that the sticking coefficients for transition metals on graphene-covered substrates are non unity and both element and temperature dependent. Our measurements of the stoichiometry by ion backscattering spectrometry (IBS) and energy dispersive spectroscopy (EDS) for films with thickness 20-80 nm provide upper bounds for the sticking coefficients of Ni and Mn on graphene/MgO, which are less than 0.6 at $600\degree$C. Controlling the stoichiometry requires compensating for the nonunity sticking coefficient on graphene, or initiating growth at low temperatures where $\sigma$ on graphene is near unity. We demonstrate epitaxial Ni$_2$MnGa films that can be mechanically exfoliated, and show how externally applied strain in Ni$_2$MnGa membranes tunes the magnetic coercive field.

\begin{figure}[h]
    \centering
    \includegraphics[width=0.45\textwidth]{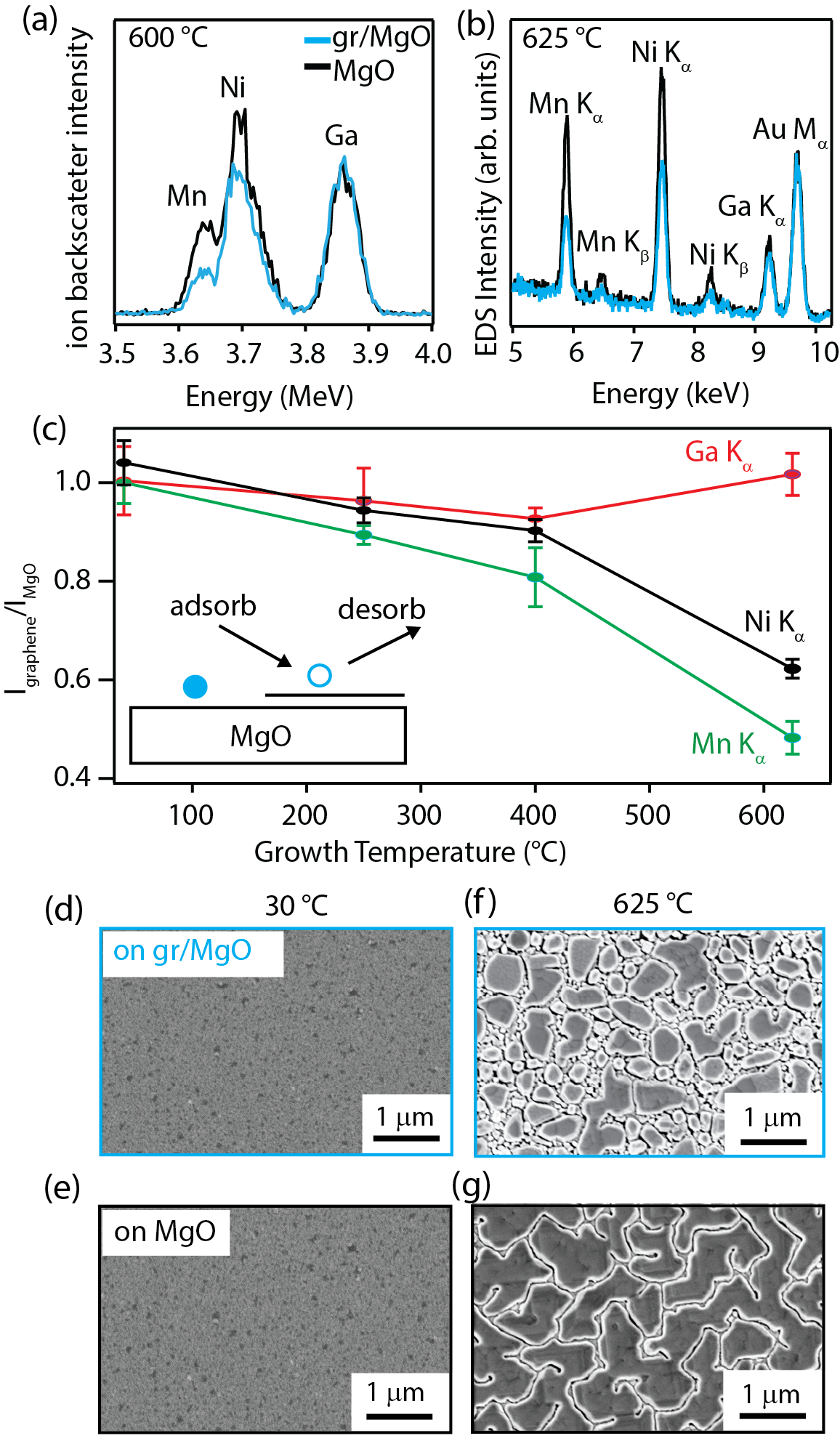}
    \caption{(a) Ion Beam Scattering (IBS) for nominally 20 nm thick Ni$_2$MnGa films grown on graphene/MgO and MgO at $600\degree$ C, showing reduced sticking for Ni and Mn on graphene. (b) Energy dispersive X-ray spectroscopy (EDS) measurements for Ni$_2$MnGa films with nominal thickness 80 nm on gr/MgO and on MgO. Both IBS and EDS sample was were capped with a protective layer of Au. 
    (c) EDS intensity ratios $I_{graphene}/I_{MgO}$, tracking temperature and element dependent changes in the cumulative sticking coefficient for Ni$_2$MnGa on graphene-covered MgO. Error bars are standard deviations on multiple regions of a given sample.
    (d,e) SEM images of the nominally 80 nm thick films grown at room remperature on graphene/MgO and MgO and capped with Au.
    (f,g) SEM images of the nominally 80 nm thick films grown at $625\degree$C on graphene/MgO and MgO and capped with Au.
    }
    \label{sticking}
\end{figure}

Ni$_2$MnGa films were grown by molecular beam epitaxy (MBE) on graphene-covered MgO (001) substrates. The graphene was grown by chemical vapor deposition on polycrystalline Cu foils and wet transferred to the MgO (001) substrate using a poly(methyl methacrylate) (PMMA) handle, Cu etch, and scoop from deionized water, as described in Refs. \cite{du2021epitaxy, du2022controlling}. Films with nominal composition Ni$_2$MnGa and nominal thickness 20-80 nm were grown by MBE using elemental effusion cell sources, with typically fluxes of $2.2 \times 10^{13}$ atoms / (cm$^2 \cdot$s) for Ni and $1.1 \times 10^{13}$ atoms / (cm$^2 \cdot$s) each for Mn and Ga. We use the term ``nominal'' to indicate the composition and thickness if all of the incident Ni, Mn, and Ga atomic fluxes stuck to the surface ($\sigma=1$). All samples were capped with $\sim 20$ nm of Au at room temperature before removal from the MBE system, to avoid oxidation.
Fluxes were measured in situ using a quartz crystal microbalance and calibrated to absolute fluxes via ex situ Ion Backscattering Spectrometry (IBS) measurements on calibration samples, grown at room temperature on Si. Energy dispersive x-ray spectroscopy (EDS, beam energy 10-20 keV, interaction depth of a few microns) was used to measure relative differences in Ni$_2$MnGa film composition. The increased depth sampling of IBS and EDS, compared to more surface sensitive XPS, allows us to sum over all species that stick, including those that may intercalate \cite{briggs2020atomically,gierz2010electronic,ebert1976intercalation} or diffuse \cite{strohbeen2021quantifying, morrow2016review} beneath graphene, rather than primarily detecting species that reside at the surface.

Fig. \ref{sticking}(a) compares IBS measurements (He$^+$, 4.9 MeV, $\theta=8\degree$) of a nominally 20 nm thick Ni$_2$MnGa film grown on graphene/MgO with a film grown directly on MgO. Growth was performed at $600 \degree$ C on an MgO substrate that is half covered with graphene, such that both sides of the sample are exposed to the same incident atomic fluxes of Ni, Mn, and Ga. We find that the areal density of Ga on graphene/MgO and on MgO are nearly equal. In contrast, the areal densities of Ni and Mn on graphene/MgO are only 50 to 60\% of the Ni and Mn on the MgO surface. Similar results are found for EDS measurements of thicker films. Fig. \ref{sticking}(b) compares EDS measurements for nominally 80 nm thick Ni$_2$MnGa films grown on graphene/MgO and MgO at $625\degree$C, where again we observe similar sticking for Ga on the graphene/MgO and MgO sides, and a $\sim 50\%$ reduction of the sticking for Ni and Mn on graphene/MgO compared to MgO.

The large stoichiometry differences in films with nominal thickness 20-80 nm are at first surprising, since differences in sticking coefficient are expected to be limited to within the first few atomic layers of growth. For planar film growth, after a layer of Ni$_2$MnGa covers the graphene or MgO surface, subsequent Ni$_2$MnGa film growth in both cases should be Ni$_2$MnGa on Ni$_2$MnGa, and thus the stoichiometries of thick films should converge. We attribute the large observed stoichiometry differences to a combination island morphology and reduced sticking coefficients on graphene. Scanning electron micrographs (SEM) reveal that the nominally 80 nm thick film grown on graphene/MgO at $625\degree$C has a disconnected island morphology (Fig. \ref{sticking}(f)), which we attribute to poor wetting on low surface energy graphene. Similar poor wetting has been observed for other films on monolayer graphene-covered surfaces \cite{strohbeen2021quantifying, alaskar2014towards}. In contrast, films grown directly on MgO at the same temperature have a smoother and more connected morphology (Fig.\ref{sticking}(g)). This morphology on graphene suggests that even after tens of nanometers of nominal growth, some exposed regions of the graphene remain. Thus our IBS and EDS measurements result from combined sticking on exposed graphene regions (where $\sigma < 1$) and on Ni$_2$MnGa islands (where $\sigma \sim 1$). This \textit{cumulative} sticking coefficient $\sigma'$ is therefor an upper bound for the true sticking coefficient $\sigma$ on graphene in the atomic layer limit. 

Importantly, we find the cumulative sticking on graphene/MgO is highly temperature and element dependent. Fig. \ref{sticking}(c) plots the EDS intensity ratio on the graphene/MgO versus on MgO ($I_{graphene}/I_{MgO}$), for a series of nominally 80 nm thick Ni$_2$MnGa films as a function of growth temperature. We normalize to the intensity on the MgO side, since the sticking coefficients for metals directly on MgO are nominally 1. Thus the ratio $I_{graphene}/I_{MgO}$ is approximately equal to the cumulative sticking coefficient on graphene/MgO. Cumulative sticking for Ni and Mn on graphene/MgO are lowest at high substrate temperature and approach 1 for all three elements below $400\degree$ C. We attribute this temperature dependence on graphene to the combined decreased desorption rate and smoother morphology with less exposed graphene at lower temperature  (Fig. \ref{sticking}(d)). We attribute the reduced sticking on graphene, compared to MgO, to relatively weak van der Waals interactions between metal adsorbates and graphene. Indeed, density functional theory calculations suggest that adsorption energies for metals on graphene are of order $E_a \sim 0.5$ eV, compared to $E_a \sim 3$ eV for Au on the Au (111) surface \cite{appy2014transition}. Surprisingly, the sticking coefficient for Ga on graphene is near unity and independent of temperature, despite the fact that Ga has a higher vapor pressure than Ni or Mn. We speculate the increased sticking of Ga on graphene may arise from reactions or from intercalation beneath graphene. Ga, In, and Sn are known to intercalate at graphene/SiC interfaces \cite{briggs2020atomically}, and Au is known to intercalate between sheets of graphite \cite{gierz2010electronic}.

Our findings suggest that control of the Ni$_2$MnGa film stoichiometry on graphene requires compensating for the non-unity sticking coefficients at high growth temperature or initiating the growth at lower temperatures where the cumulative sticking coefficients are closer to 1.  Lower temperature growth is also beneficial for promoting a smoother morphology on graphene (Fig. \ref{sticking}(d)). Once the interface has formed and the graphene layer is buried, growth can resume under more normal temperatures and fluxes. 

\begin{figure}[h]
    \centering
    \includegraphics[width=0.45\textwidth]{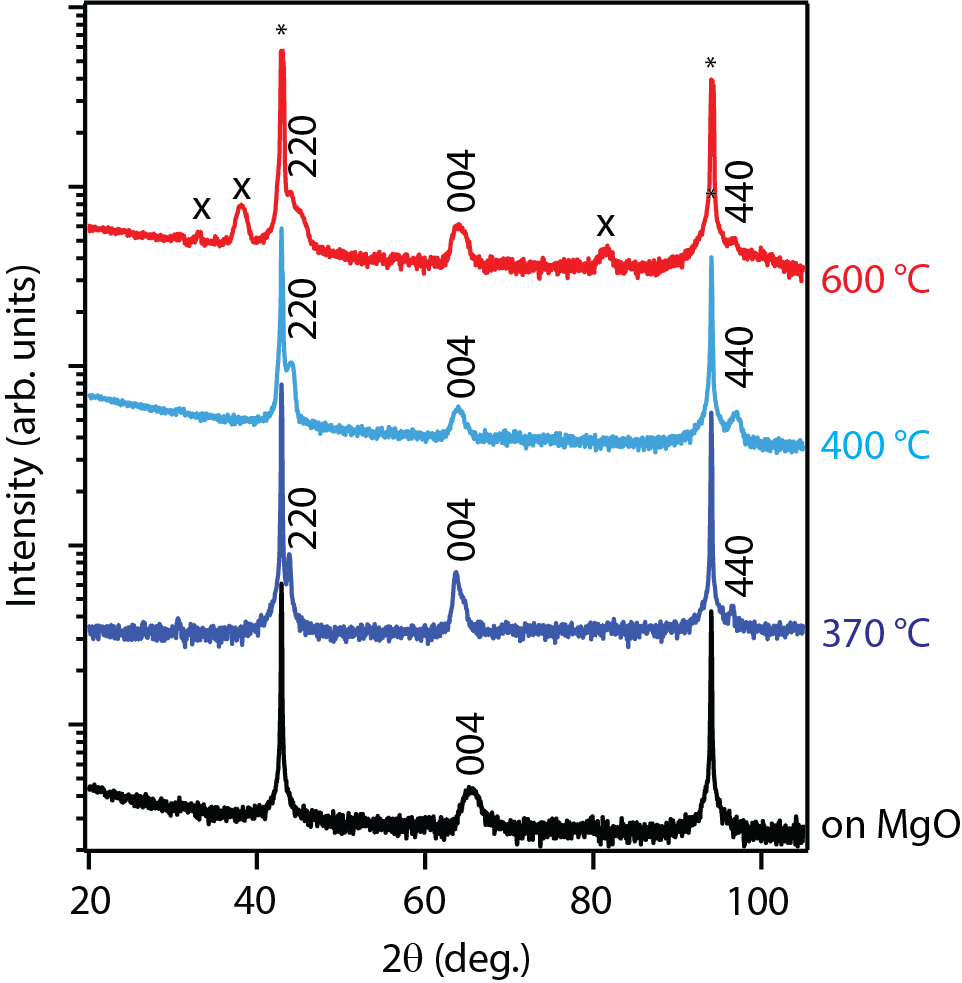}
    \caption{Out of plane X-ray diffraction scans (Cu $K\alpha$) of Ni$_2$MnGa films grown on MgO and on graphene/MgO at $600\degree$C, $400\degree$C, and $370\degree$C, compared to a film grown directly on MgO. Asterisks * denote MgO substrate reflections and ``x'' denotes secondary phase reflections.}
    \label{synthesis}
\end{figure}

For simplicity adopt the strategy of growth at a fixed lower temperature. Fig. \ref{synthesis} compares x-ray diffraction patterns for Ni$_2$MnGa films grown at $370\degree$ C, $400\degree$ C, and $600\degree$ C on graphene/MgO, with a film grown by directly on MgO. We find that the sample grown at $600\degree$ C on graphene/MgO displays several impurity reflections, marked by ``x,'' consistent with large deviations from stoichiometry observed by IBS and EDS (Fig. \ref{sticking}). Films grown at $400 \degree$C and below on graphene display only the Heusler Ni$_2$MnGa reflections and no impurity reflections. 
Interestingly, the films on graphene/MgO display both $00L$ and $HH0$ reflections, indicating both (001) and (110) oriented growth, whereas epitaxy directly on MgO produces only (001) oriented growth (black curve). 

\begin{figure}[ht]
    \centering
    \includegraphics[width=0.45\textwidth]{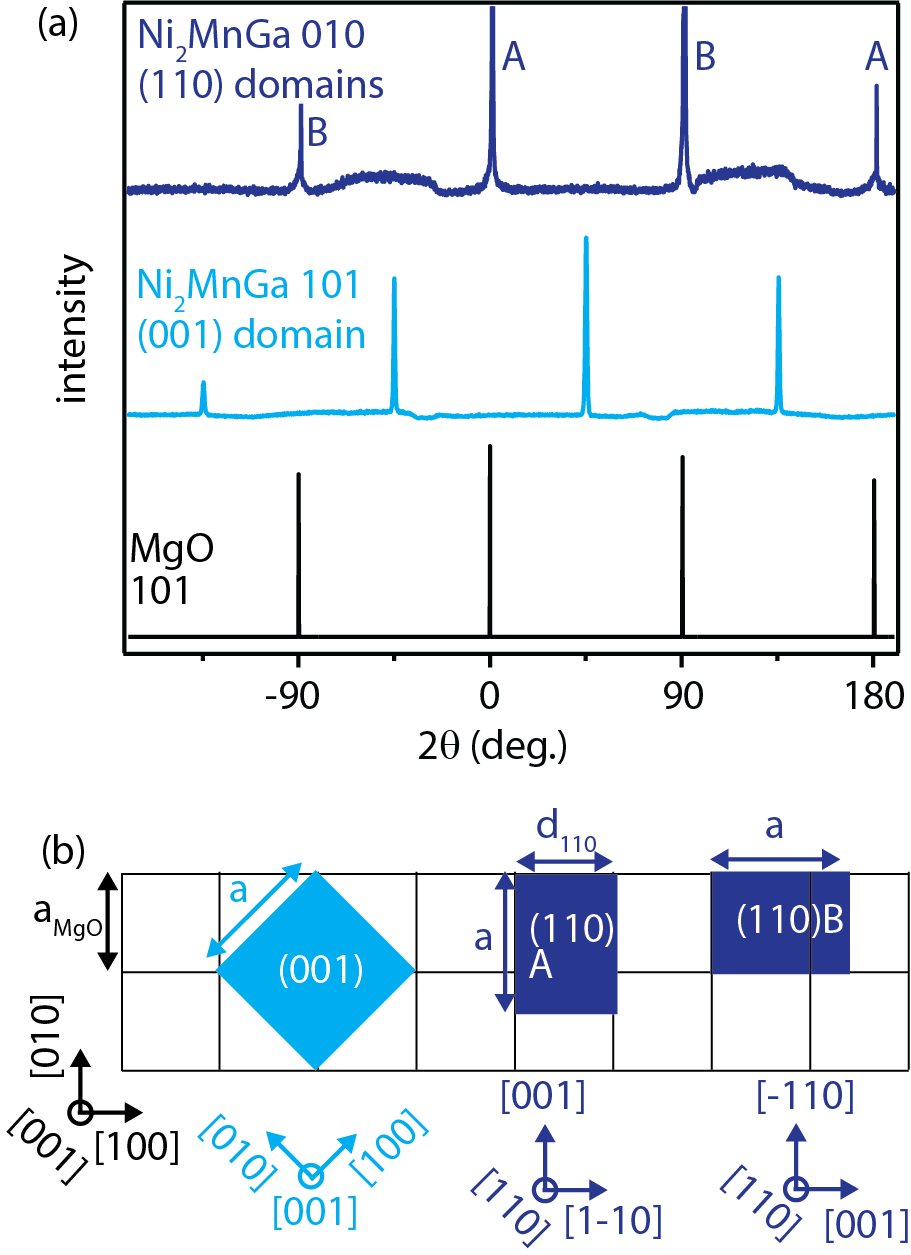}
    \caption{(a) Azimuthal $\phi$ scans for a Ni$_2$MnGa film grown on graphene/MgO (001). The off axis $010$ reflections track the in-plane orientation of Ni$_2$MnGa domains with (110) out of plane orientation. The $101$ reflections track the in-plane orientation of the (001) domain. (b) Domain orientations of Ni$_2$MnGa (blue) with respect to MgO (001) (black) determined from (a).}
    \label{pole}
\end{figure}

Azimuthal $\phi$ scans reveal that both (110) and (001) oriented Ni$_2$MnGa domains on graphene/MgO (001) have well defined in-plane orientations with respect to the underlying MgO substrate, despite the presence of the polycrystalline graphene interlayer. In Fig. \ref{pole}(a), the four-fold pattern of Ni$_2$MnGa 101 reflections is rotated by 45 degrees with respect to the MgO 101. This indicates that the (001) Ni$_2$MnGa domain has a 45 degree rotated cube on cube epitaxial relationship to MgO, i.e. Ni$_2$MnGa (001) [110] $\parallel$ MgO (001) [100], and a $2\%$ tensile lattice mismatch (Fig. \ref{pole}(b)). For the (110) domain, we observe a four-fold pattern of 010 Ni$_2$MnGa reflections aligned with the MgO 101. This indicates two rectangular domains, labelled A and B, with orientations Ni$_2$MnGa (110) [001] $\parallel$ MgO (001) [010] and Ni$_2$MnGa (110) [001] $\parallel$ MgO (001) [100] (Fig. \ref{pole}(b)). For these (110)-orientated domains, the mismatch between Ni$_2$MnGa $d_{110}$ and the MgO $a$ lattice spacings is $2\%$, while the mismatch in the orthogonal in-plane direction ($a_{Ni2MnGa}$ vs $a_{MgO}$) is much larger and would require a larger supercell to produce a commensurate structure. We speculate that the presence of the graphene interlayer may relax the constraints of direct epitaxy in which there are direct bonds formed between film and substrate, and allow for alternative film orientations that lower the total energy. Similar new epitaxial structures have been observed in the form of rotated superstructures for GdPtSb films on graphene/sapphire \cite{du2022controlling}. Further studies are required to understand why the (110) domain appears on graphene/MgO and not directly on MgO.


\begin{figure}[h]
    \centering
    \includegraphics[width=0.45\textwidth]{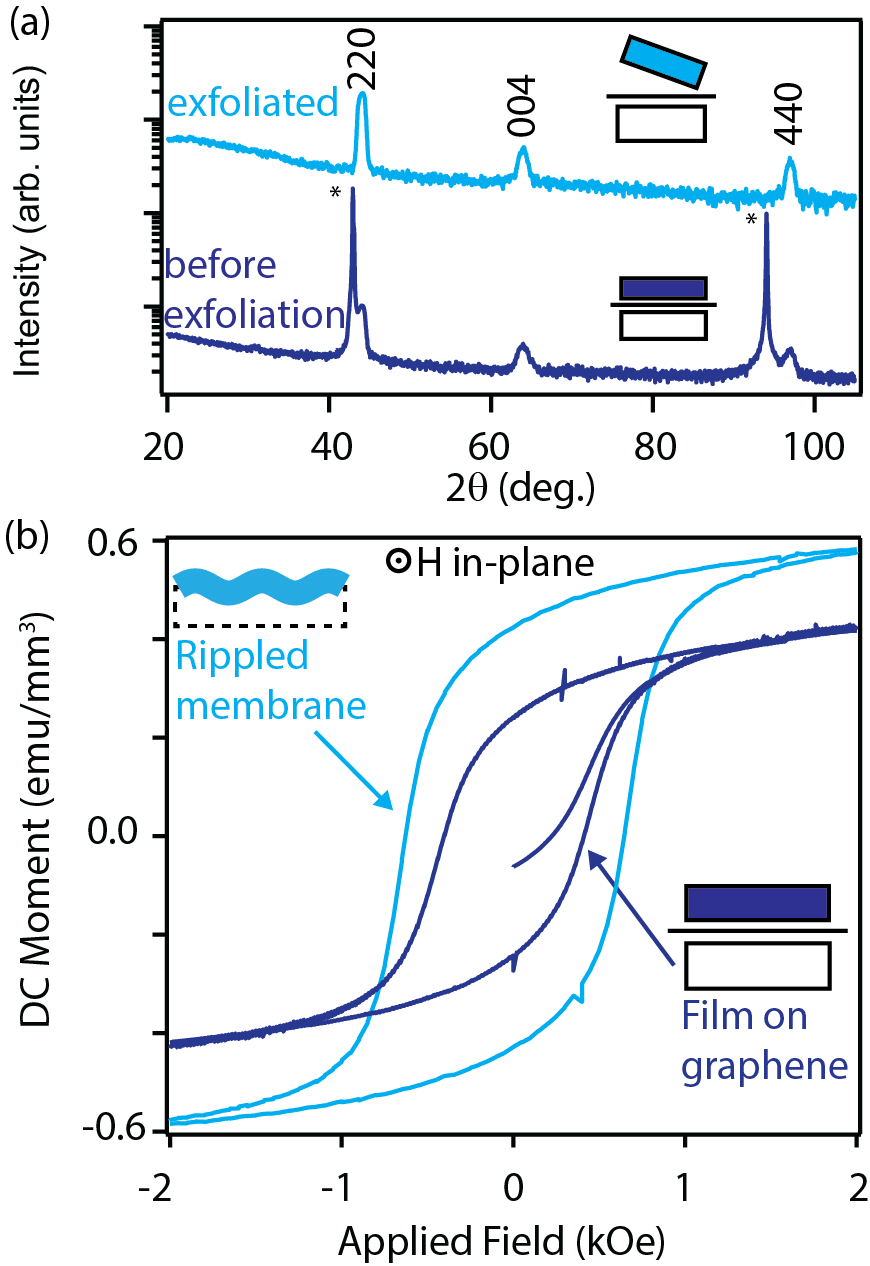}
    \caption{(a) X-ray diffraction before and after membrane exfoliation.
    (b) SQUID magnetometry of a relaxed Ni$_2$MnGa film on graphene/MgO (dark blue), and on the same sample after exfoliation and rippling to create a strained Ni$_2$MnGa membrane (light blue). The measurement was performed at 100 K with field oriented within the film plane.}
    \label{magnetization}
\end{figure}

Finally, we show that applying external strains to exfoliated membranes tunes magnetic properties. We exfoliate membranes by  adhering the film to a glass slide using crystal bond, then peeling the film from the graphene/MgO. After exfoliation we observe only the 110 and 001-type film reflections and no substrate reflections, as shown in Fig. \ref{magnetization}(a). We then apply ripples to the membrane to induce strain. The rippling was performed by adhering a tensile strained polyurethane film to the exfoliated Ni$_2$MnGa membrane, heating to approximately $150\degree$C to release the Ni$_2$MnGa/polyurethane bilayer from the crystalbond, and relaxing to impart ripples upon contraction of the polyurethane. Further details of the rippling procedure are described in Ref. \cite{du2021epitaxy}.

Fig. \ref{magnetization}(b) shows SQUID magnetometry measurements for a Ni$_2$MnGa film on graphene/MgO, and the same film after exfoliation and subjected to strain in the form of rippling, measured at 100 K  with field oriented in plane.
We find that strain and/or strain gradients enhance the coercive field, from 400 to 650 Oe. The membrane has thickness 80 nm, ripple period of 8 microns, and peak to peak height of 3 microns. Assuming a sinusoidal shape we we estimate the peak magnitudes of strain to be $|\epsilon |< 3.6\%$, if no plastic deformation \cite{du2021epitaxy}. 
The strain tunable coercive field may be useful for strain-assisted reading and writing of magnetic memory.

In summary, we showed that sticking coefficients for metals on graphene-covered substrates are non unity and highly dependent on element and temperature. IBS and EDS measurements of films with tens of nanometers thickness provide upper bounds for the sticking coefficients on graphene/MgO: $\sigma<0.6$ for Ni and Mn and $\sigma<1$ for Ga at $600\degree$C. Surface sensitive measurements in the monolayer limit are required to fully quantify the atomic sticking coefficients on graphene, and understand the effects of changing the underlying substrate and effects of defects and contaminants at the graphene/substrate interface. In particular, the lattice potential permeation argument of remote epitaxy\cite{kim2014principle, kong2018polarity} suggests that the sticking coefficients should also depend on the identity of the substrate. We show that synthesis at lower temperature $\leq 400 \degree$C enables phase pure epitaxy of Ni$_2$MnGa films on graphene/MgO. Similar strategies may apply to remote and van der Waals epitaxy of other materials with complex stoichiometries, for which adsorption-controlled growth windows are not accessible.


\section*{Acknowledgment}

We thank Greg Haugsted for IBS/RBS measurements. This work was primarily supported by the Air Force Office of Scientific Research grant FA9550-21-0127 (Z.L. and J.K.K.). Preliminary Heusler synthesis was supported by the National Science Foundation DMR-1752797 (Z.L., S.M., and J.K.K.). Graphene synthesis via CVD supported by the U.S. Department of Energy, Office of Science, Basic Energy Sciences, Grant DE-SC0016007 (K.S. and M.S.A.).

We gratefully acknowledge the use of x-ray diffraction facilities supported by the NSF through the University of Wisconsin Materials Research Science and Engineering Center under Grant No. DMR-1720415.

\bibliographystyle{apsrev}
\bibliography{ref}
\end{document}